\documentclass[preprint, authoryear]{elsarticle}
\usepackage{soul, color}
\usepackage{colortbl}
\usepackage{bm}
\usepackage{graphicx}
\usepackage{booktabs}
\usepackage{tikz}
\usepackage{caption}
\usepackage{arydshln}
\usepackage{subcaption}
\usepackage{amsmath}
\usepackage{amssymb}
\usepackage{amsthm}
\usepackage{dsfont}
\usepackage{bbold}
\usepackage{listings}
\usepackage{setspace}
\usepackage{url}
\usepackage{verbatim}
\usepackage{dsfont}
\usepackage{float}
\usetikzlibrary{positioning,decorations.pathreplacing,shapes}

\newcommand{\E}{\mathbb{E}}
\newcommand{\Var}{\textrm{Var}}
\newcommand{\prob}{Prob}

\newcommand{\uT}{\widetilde{\mathbf{u}}}
\newcommand{\bT}{\widetilde{\mathbf{b}}}

\newcommand{\qT}{\widetilde{\mathbf{q}}}
\newcommand{\BT}{\widetilde{\mathbf{B}}}
\newcommand{\UT}{\widetilde{\mathbf{U}}}
\newcommand{\SigT}{\widetilde{\mathbf{\Sigma}}}

\newcommand{\yH}{\widehat{\mathbf{y}}}
\newcommand{\bH}{\widehat{\mathbf{b}}}
\newcommand{\uH}{\widehat{\mathbf{u}}}

\newcommand{\qH}{\widehat{\mathbf{q}}}
\newcommand{\BH}{\widehat{\mathbf{B}}}
\newcommand{\UH}{\widehat{\mathbf{U}}}
\newcommand{\YH}{\widehat{\mathbf{Y}}}
\newcommand{\SigH}{\widehat{\mathbf{\Sigma}}}

\newcommand{\pitil}{\widetilde{\pi}}
\newcommand{\ptil}{\widetilde{p}}

\newcommand{\util}{\widetilde{u}}

\newcommand{\Btil}{\widetilde{B}}
\newcommand{\Util}{\widetilde{U}}

\newcommand{\Bhat}{\widehat{B}}
\newcommand{\Uhat}{\widehat{U}}

\newcommand{\uhat}{\widehat{u}}
\newcommand{\phat}{\widehat{p}}
\newcommand{\qhat}{\widehat{q}}
\newcommand{\yhat}{\widehat{y}}
\newcommand{\sigmahat}{\widehat{\sigma}}
\newcommand{\pihat}{\widehat{\pi}}

\newcommand{\lambdahat}{\widehat{\lambda}}

\newcommand{\Bbf}{\mathbf{B}}

\newcommand{\Sbf}{\mathbf{S}}
\newcommand{\Gbf}{\mathbf{G}}
\newcommand{\Ibf}{\mathbf{I}}
\newcommand{\Abf}{\mathbf{A}}

\newcommand{\Hbf}{\mathbf{H}}
\newcommand{\Zbf}{\mathbf{Z}}
\newcommand{\Qbf}{\mathbf{Q}}
\newcommand{\Tbf}{\mathbf{T}}

\newcommand{\Dbf}{\mathbf{D}}
\newcommand{\Mbf}{\mathbf{M}}
\newcommand{\Cbf}{\mathbf{C}}
\newcommand{\Ebf}{\mathbf{E}}

\newcommand{\bbf}{\mathbf{b}}
\newcommand{\ubf}{\mathbf{u}}
\newcommand{\ybf}{\mathbf{y}}
\newcommand{\xbf}{\mathbf{x}}

\newcommand{\sbf}{\mathbf{s}}

\newcommand{\alphabf}{\bm{\alpha}}

\newcommand{\mubf}{\bm{\mu}}

\newcommand{\epsbf}{\mathbf{\epsilon}}

\newcommand{\rr}{\mathbb{R}}

\newcommand{\Poi}{Poi}

\definecolor{codegreen}{rgb}{0,0.6,0}

\definecolor{codegray}{rgb}{0.5,0.5,0.5}
\definecolor{codepurple}{rgb}{0.58,0,0.82}
\definecolor{backcolour}{rgb}{0.95,0.95,0.92}
\definecolor{Gray}{gray}{0.8}
\definecolor{lightyellow}{RGB}{245,238,197}
\newcolumntype{a}{>{\columncolor{lightyellow}}c}
\newcolumntype{b}{>{\columncolor{backcolour}}c}

\lstdefinestyle{mystyle}{
    backgroundcolor=\color{backcolour},
    commentstyle=\color{codegreen},
    keywordstyle=\color{magenta},
    numberstyle=\tiny\color{codegray},
    stringstyle=\color{codepurple},
    basicstyle=\ttfamily\footnotesize,
    breakatwhitespace=false,
    breaklines=true,
    captionpos=b,
    keepspaces=true,
    numbers=left,
    numbersep=5pt,
    showspaces=false,
    showstringspaces=false,
    showtabs=false,
    tabsize=2
}

\DeclareFontFamily{U}{mathx}{}
\DeclareFontShape{U}{mathx}{m}{n}{<-> mathx10}{}
\DeclareSymbolFont{mathx}{U}{mathx}{m}{n}
\DeclareMathAccent{\widehat}{0}{mathx}{"70}
\DeclareMathAccent{\widecheck}{0}{mathx}{"71}

\newtheorem{proposition}{Proposition}

\begin{document}
\begin{frontmatter}
\title{Properties of the reconciled distributions for Gaussian and count forecasts}

\author[1,2]{Lorenzo Zambon\corref{cor1}}
\ead{lorenzo.zambon@idsia.ch}

\author[2]{Arianna Agosto}
\ead{arianna.agosto@unipv.it}

\author[2]{Paolo Giudici}
\ead{paolo.giudici@unipv.it}

\author[1]{Giorgio Corani}
\ead{giorgio.corani@idsia.ch}

\cortext[cor1]{Corresponding author}

\affiliation[1]{organization={Istituto Dalle Molle di Studi sull'Intelligenza Artificiale (IDSIA)},
city={Lugano},
country={Switzerland}}

\affiliation[2]{organization={University of Pavia},
country={Italy}}

\begin{abstract}
Reconciliation enforces coherence between hierarchical forecasts, in order to satisfy a set of linear constraints.
While most works focus on the reconciliation of the point forecasts, we consider probabilistic reconciliation and
we  analyze the properties of the distributions  reconciled via conditioning. 
We provide a formal analysis of the variance of the reconciled distribution,
treating separately  the 
case of Gaussian forecasts and count forecasts. 
We also study  the reconciled upper mean in the case of 1-level hierarchies; also in this case we analyze separately the case of Gaussian forecasts and count forecasts.
We then show experiments on the reconciliation of intermittent time series related to the count of extreme market events. 
The experiments confirm our theoretical results and show that reconciliation largely improves the performance of probabilistic forecasting.
\end{abstract}

\end{frontmatter}

\section{Introduction}

Hierarchical forecasting requires  the forecasts to be coherent, i.e., to  satisfy a set of linear constraints determined by the  structure of the hierarchy.
The base forecasts, computed  independently on each time series of the hierarchy are incoherent; \textit{reconciliation} adjusts them 
to enforce coherence.

Most  reconciliation approaches  reconcile the point forecasts  \citep{hyndman2011optimal, Wickramasuriya.etal2018, PANAGIOTELIS2021343, difonzo2022forecast}.
However, \textit{reconciled predictive distributions} are required to support decision making \citep{kolassa22we}.
\cite{panagiotelis2023probabilistic} performs
probabilistic reconciliation through projection,
learning the parameters of the projection via stochastic gradient descent.
Two limits of this approach is that  it cannot reconcile count variables and it prevents an analytical study of  the reconciled  distribution. These considerations are valid also  for other methods for  probabilistic reconciliation \citep{jeon2019probabilistic, taieb2021hierarchical, rangapuram2021end, hollyman2022hierarchies}.

A different approach to probabilistic forecast reconciliation is constituted
by  reconciliation via conditioning.
In the case of Gaussian base forecasts, it yields the reconciled distribution in closed form  \citep{corani_reconc}, with the same mean and variance of MinT \citep{Wickramasuriya.etal2018}.
Reconciliation via conditioning can also reconcile the distribution of count variables, 
adopting a sampling approach  \citep{corani2023, zambon2022}.
 
In this paper, we study the properties of the  reconciled distribution obtained via conditioning.
In the  Gaussian case we  prove that, regardless the amount of  incoherence of the base forecasts, reconciliation  decreases
the variance of every variable of the hierarchy.
In contrast, we show that in the discrete case reconciliation con increase or decrease
the variance of the bottom variables, depending on  the probability $p_c$ of the base  forecasts being coherent.

We then analyze the reconciled mean,  restricting  our analysis to the case of 1-level hierarchies. In the Gaussian case 
the reconciled upper mean is a combination of the bottom-up mean and the mean of the upper base forecast \citep{corani_reconc, hollyman2021understanding};
we refer to this as the \textit{combination}  effect.
However, in the case of count variables  we  show that
the reconciled mean  of the upper time series can be lower than both the bottom-up  and the base mean:
we refer to this as  the \textit{concordant-shift} effect,
as the reconciled means of all the time series are shifted towards zero.
This happens when the base forecast distributions are right-skewed and reconciliation decreases the  variance of the forecasts, shortening the  right tails of the distributions and pulling the reconciled means of all time series towards zero.
In other words, low counts forecasts across the hierarchy reinforce each other.

We present experiments on the reconciliation of  intermittent time series referring to counts of extreme market events, interpreting them on the basis of  our theoretical insights. 
We show that, on the intermittent time series of our case study,  
the concordant-shift effect  on the mean is more common than the combination effect.
We moreover report a major increase of  accuracy for the probabilistic forecasts after reconciliation, confirming  the beneficial effect of reconciliation for
forecasting intermittent time series \citep{athanasopoulos2017_temporal, KOURENTZES-elucidate,
 corani2023, zambon2022}.

The paper is organized as follows. 
In Section~\ref{sec: prob fore rec}, we recall probabilistic reconciliation through conditioning, and analyze the reconciled mean and variance of the reconciled distribution in the Gaussian and in the non-Gaussian case.
In Section~\ref{sec:experiments}, we present our case study.
Finally, the conclusions are in Section~\ref{sec: conclusions}.

\section{Probabilistic forecast reconciliation}
\label{sec: prob fore rec}
Given a hierarchy,
we denote by $\bbf = [b_1,\dots,b_{n_b}]^T$ the vector of bottom variables, and by $\ubf = [u_1,\dots,u_{n_u}]^T$ the vector of upper variables. 
To keep the notation simple, we do not show the time index; it is thus understood that  all forecasts refer to the time $t+h$. 
We denote the vector of all the variables by
\[
\ybf = \begin{bmatrix}
          \ubf \\
          \bbf
         \end{bmatrix} \in \rr^n.
\]
The hierarchy may be expressed as a set of linear constraints:
\begin{equation}\label{eq: def S}
\ybf = \Sbf \bbf, \quad \text{with} \;\; \Sbf= \begin{bmatrix}
          \Abf \\ \hdashline[2pt/2pt]
          \Ibf
         \end{bmatrix},
\end{equation}
where $\Ibf \in \rr^{n_b \times n_b}$ is the identity matrix. 
$\Sbf \in \rr^{n \times n_b}$ is the \textit{summing matrix}, while $\Abf \in \rr^{n_u \times n_b}$ is the \textit{aggregating matrix}. 
The summing constraints can thus be written as $\ubf = \Abf \bbf$.

We assume the base forecasts to be in the form of predictive distributions.
We denote by $\pihat$ the base forecast distribution for the entire hierarchy,
and by $\pihat_U$ and $\pihat_B$ the  base forecasts for the upper and the bottom variables.
The aim of probabilistic reconciliation is to find a reconciled distribution $\pitil$ that gives positive probability only to coherent points.
To this end, we first obtain  a reconciled bottom distribution $\pitil_B$ from the base forecast distribution $\pihat$.
Then, we obtain the reconciled distribution $\pitil$ on the entire hierarchy as: 
\[
\pitil(\ubf, \bbf) =
\begin{cases}
\pitil_B(\bbf) \quad \quad &\text{if } \ubf = \Abf \bbf \\
0  &\text{if } \ubf \neq \Abf \bbf,
\end{cases}
\]
so that the probability of any set of incoherent points is zero.

\paragraph{Probabilistic bottom-up}
If we set $\pitil_B = \pihat_B$, we have the probabilistic bottom-up, which ignores the base forecast distribution $\pihat_U$ of the upper variables of the hierarchy.
The reconciled bottom-up distribution $\pitil_{bu}$ is thus given by:
\begin{equation}\label{eq: probabilistic bottom up}
 \pitil_{bu}(\ubf, \bbf) =
\begin{cases}
\pihat_B(\bbf) \quad \quad &\text{if } \ubf = \Abf \bbf \\
0  &\text{if } \ubf \neq \Abf \bbf.
\end{cases}   
\end{equation}

\paragraph{Reconciliation through conditioning}
In this work, we apply reconciliation via conditioning.
In order to  take into account  the base forecasts of all variables,
we introduce the random vector
\[
\YH = \begin{bmatrix} \UH \\ \BH \end{bmatrix} \sim \pihat,
\]
so that $\pihat_U$ and $\pihat_B$ are the distributions of  $\UH$ and $\BH$. 
We then define the reconciled bottom distribution by conditioning on the hierarchy constraints.
For discrete distributions  we have:
\begin{align}
\pitil_B(\bbf) &:=
\prob\big(\BH = \bbf \mid \UH - \Abf \BH = 0 \big) \nonumber\\
&= \frac{\prob\big(\BH = \bbf, \; \UH - \Abf \BH = 0 \big)}{\prob\big(\UH - \Abf \BH = 0 \big)} \nonumber \\
&= \frac{\prob\big(\BH = \bbf, \; \UH = \Abf \bbf \big)}
{\sum_{\xbf \in \rr^m} \prob\big(\BH = \xbf, \; \UH = \Abf \xbf \big)} \nonumber \\
&= \frac{\pihat(\Abf \bbf,\bbf)}{\sum_{\xbf \in \rr^m} \pihat(\Abf \xbf,\xbf)} \nonumber \\
&\propto \pihat(\Abf \bbf,\bbf). \label{eq: reconciled distribution}
\end{align}
The same formula, $\pitil_B(\bbf)\propto \pihat(\Abf \bbf,\bbf)$, 
also holds for continuous distributions.
We refer to \cite{zambon2022} for the derivation in the continuous case.

\subsection{Gaussian reconciliation}
\label{sec: gaussian case}
In the Gaussian case, reconciliation via conditioning can be solved in closed form
\citep{corani_reconc}; its  reconciled mean and variance are numerically equivalent to those of  MinT \citep{Wickramasuriya.etal2018}, despite the different derivation. 

In \ref{appendix: derivation gaussian} we derive in a novel way the reconciliation  formulae;  they are equivalent to those of \cite{corani_reconc} and \cite{Wickramasuriya.etal2018},  but more convenient to prove some
properties.
Let us assume the base forecasts for the entire hierarchy to be multivariate Gaussian:
\begin{equation}
\YH = \begin{bmatrix} \UH \\ \BH \end{bmatrix}
\sim \mathcal{N}\left( \yH, \,\SigH_Y \right),
\end{equation}
where
\begin{equation*}
\yH = \begin{bmatrix} \uH \\ \bH \end{bmatrix}, \quad
\SigH_Y = \begin{bmatrix} \SigH_U & \SigH_{UB} \\
\SigH_{UB}^T & \SigH_B \end{bmatrix}.
\end{equation*}
In \ref{appendix: derivation gaussian}, we show that this is equivalent to the framework of \cite{corani_reconc}.
The covariance matrix of the base forecasts $\SigH_Y$ can thus be computed in practice as the covariance matrix of the forecasting errors.
Assuming $\SigH_Y$ to be positive definite, the reconciled bottom and upper distributions are  multivariate Gaussian:
\begin{equation*}
\BT \sim \mathcal{N}\left(\bT, \,\SigT_B \right), \qquad
\UT \sim \mathcal{N}\left(\uT, \,\SigT_U \right),
\end{equation*}
where
\begin{align}
\bT &= \bH + \left(\SigH_{UB}^T - \SigH_B \Abf^T\right) \Qbf^{-1} (\Abf \bH - \uH), \label{eq: reconciled gaussian mean bottom} \\
\uT &= \uH + \left(\SigH_U - \SigH_{UB} \Abf^T\right) \Qbf^{-1} (\Abf \bH - \uH), \label{eq: reconciled gaussian mean upper} \\
\SigT_B &= \SigH_B - \left(\SigH_{UB}^T - \SigH_B \Abf^T\right) \Qbf^{-1} \left(\SigH_{UB}^T - \SigH_B \Abf^T\right)^T, \label{eq: reconciled gaussian variance bottom} \\
\SigT_U &= \SigH_U - \left(\SigH_U - \SigH_{UB} \Abf^T\right) \Qbf^{-1} \left(\SigH_U - \SigH_{UB} \Abf^T\right)^T, \label{eq: reconciled gaussian variance upper}
\end{align}
and $\Qbf:= \SigH_U - \SigH_{UB} \Abf^T - \Abf \SigH_{UB}^T + \Abf \SigH_B \Abf^T$.

\begin{proposition}[Reconciled Gaussian variance]\label{prop: var gauss}
In the Gaussian framework, the variance of each variable decreases after reconciliation.\\
Indeed, for each $i=1,\dots,m$, and $j=1,\dots,n-m$:
\begin{align}\label{eq: variance inequality prop}
\Var(\Btil_i) \le \Var(\Bhat_i), \nonumber \\
\Var(\Util_j) \le \Var(\Uhat_j).
\end{align}
\end{proposition}
The proof is given in \ref{app: proof var ineq}. 
We remark that the  variance of each variable decreases after reconciliation, regardless of the amount of incoherence.
While we consider  reconciliation via conditioning,
\cite{wickramasuriya2021properties} provides  a similar result for the minT reconciliation.
In \cite{wickramasuriya2021properties}, no Gaussian assumption is made; 
however, it assumes the unbiasedness of the base forecast, which we do not assume.

\paragraph{Reconciled Gaussian mean}
In order to study the reconciled mean, we need some 
restrictive assumptions (which, however, do apply to the  case study of  Sect.~\ref{sec:experiments}).
In particular, assuming that:
\begin{itemize}
    \item there is only one upper variable,
    \item  there is no correlation between the bottom and the upper base forecasts,
\end{itemize}
the reconciled upper mean is a convex combination of  $\uhat$ and  $\Abf \bH$.
Indeed, from \eqref{eq: reconciled gaussian mean upper}, we have:
\begin{equation}\label{eq: gauss reconciled mean upper}
\util = \frac{\sigmahat^2_{bu}}{\sigmahat^2_U + \sigmahat^2_{bu}} \, \uhat + \frac{\sigmahat^2_U}{\sigmahat^2_U + \sigmahat^2_{bu}} \, \Abf \bH,
\end{equation}
where  $\sigmahat^2_U$ is the variance of the upper base forecast and $\sigmahat^2_{bu} := \Abf \SigH_B \Abf^T \ge 0$ is the variance of the probabilistic bottom-up, defined in \eqref{eq: probabilistic bottom up}.
The reconciled mean is thus a combination of the base and the bottom-up mean, 
as already observed \citep{corani_reconc, hollyman2021understanding}.
We call this the \textit{combination effect}.

We can draw an analogy with  the Gaussian conjugate model in Bayesian statistics: the posterior variance is always smaller than the prior variance \citep{Gelman2011},
and the posterior expectation is a convex combination of the prior expectation and the sample mean \citep[Ch.~2.5]{bda2013}.


\subsection{Sampling from the reconciled distribution}
\label{sec: sampling from the rec distr}
In the non-Gaussian  case,  the reconciled distribution $\pitil$  
is in general not available in a parametric form; 
hence, we need to draw samples from $\pitil$.
We follow the approach of  \cite{zambon2022} based on importance sampling \citep{elvira2021advances}.

Denoting by $N$ the number of samples, the algorithm is as follows:
\begin{enumerate}
    \item Sample $\big(\bH^{(i)}\big)_{i=1,\dots,N}$ from $\pihat_B$
    \item Compute the unnormalized weights $\big(\widecheck{w}^{(i)}\big)_{i=1,\dots,N}$ as 
    \[\widecheck{w}^{(i)} = \frac{\pihat\big(\Abf \bH^{(i)}, \bH^{(i)}\big)}{\pihat_B\big(\bH^{(i)}\big)}.\]
    If we assume the  bottom and upper base forecasts to be independent, as we do in this paper, we have
    \[\widecheck{w}^{(i)} = \frac{\pihat_U\big(\Abf \bH^{(i)}\big) \pihat_B\big(\bH^{(i)}\big)}{\pihat_B\big(\bH^{(i)}\big)}
    = \pihat_U\big(\Abf \bH^{(i)}\big)\]
    \item Compute the normalized weigths as $w^{(i)} = \widecheck{w}^{(i)} / \sum_h \widecheck{w}^{(h)}$
    \item Sample $\big(\bT^{(i)}\big)_i$ with replacement from the weighted sample $\big( \bH^{(i)}, w^{(i)} \big)_{i=1,\dots,N}$
\end{enumerate}
The output $\big(\bT^{(i)}\big)_i$ is an unweighted sample from the reconciled distribution $\pitil_B$.
The algorithm  assumes the base forecasts of the upper and the bottom variables to be conditionally independent, i.e., to be independent given the observations available up to time $t$.

Computationally, such algorithm is suitable for the reconciliation of small hierarchies  such those considered in this paper (a single upper variable). 
Larger hierarchies can be reconciled by an extension of this algorithm, called bottom-up importance sampling (BUIS, \cite{zambon2022}).
The BUIS algorithm is implemented in the \textit{R} package \textit{bayesRecon} \citep{bayesRecon}.

\subsection{Reconciled variance of discrete variables}
\label{sec: rec var}

In the case of \textit{discrete} variables, the variance of the reconciled distribution
of the bottom variables can be larger than the variance of the base distribution;
this is  a major difference with the Gaussian reconciliation.
In particular, this happens if the  probability $p_c := \prob\big(\UH=\Abf\BH\big)$
of the base forecasts  being coherent is small.

\begin{proposition} 
\label{prop: reconc variance}
Let us assume  $\BH$ and $\UH$ to be discrete random variables with $p_c > 0$.
Then, for any $j=1,\dots,m$:
\begin{equation}
\label{eq: reconciled variance}
\Var\big[\Btil_j\big]
= \frac{\Var\big(\Bhat_j\big) - (1-p_c) \, \Var\big[\Bhat_j \,|\, \UH \neq \Abf \BH\big] 
- p_c (1-p_c) \, (a-b)^2}{p_c},
\end{equation}
where $a := \E\big[\Bhat_j \,|\, \UH \neq \Abf \BH \big]$ and $b := \E\big[\Bhat_j \,|\, \UH = \Abf \BH \big]$. 
\end{proposition}
The proof is  in  \ref{app: proof rec var}.
According to Eq. \eqref{eq: reconciled variance}, a small $p_c$ can result in a large variance of the reconciled bottom distributions.
We can draw yet another  analogy with Bayesian statistics:
in non-Gaussian models, the posterior variance can be larger than the prior variance if we condition on observations that are conflicting with the prior beliefs \citetext{\citealp[Ch.~2.2]{bda2013}; \citealp{Gelman2011}}.

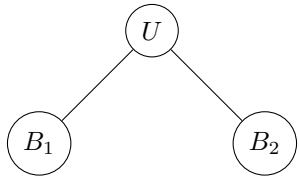
\begin{figure}[!h]
    \centering
\begin{tikzpicture}[level/.style={sibling distance=30mm/#1}]
\node [circle,draw] {$U$}
  child {node [circle,draw]  {$B_1$}
  }
  child {node [circle,draw]  {$B_2$}
};
\end{tikzpicture}
 \caption{\label{fig: simple hier gauss} A minimal hierarchy.}
\end{figure}

\paragraph*{Reconciling a minimal hierarchy}
We now illustrate  the increase of variance after reconciliation on the minimal hierarchy  of Fig.~\ref{fig: simple hier gauss}.
Consider the following independent base forecasts:
\begin{align*}
\Bhat_1 \sim &\text{Bernoulli}\,(\phat_1), \quad
\Bhat_2 \sim \text{Bernoulli}\,(\phat_2), \\
& \quad \Uhat = \begin{cases}
0 \qquad \text{prob} = \qhat_0 \\
1 \qquad \text{prob} = \qhat_1 \\
2 \qquad \text{prob} = \qhat_2,
\end{cases} 
\end{align*}
where $\phat_1,\phat_2 \in [0,1]$, $\qhat_0, \qhat_1, \qhat_2 \in [0,1]$, and $\qhat_0 + \qhat_1 + \qhat_2 = 1$.
In \ref{app: example bernoulli}, we analytically derive the expression of the parameters $\ptil_1, \ptil_2$, and $\qT$ of the reconciled distribution.

\begin{table}[!ht]
    \centering
    \small
\begin{tabular}{c @{\hskip 8mm} c c @{\hskip 8mm} c c c}
\toprule
   &  \multicolumn{2}{c@{\hspace{8mm}}}{\textbf{mean}} & \multicolumn{3}{c}{\textbf{variance}}\\
   & \textit{base} & \textit{reconc} & \textit{base} & \textit{reconc}
   & $\Delta$\\
\midrule
\textit{Bernoulli} & & & & &\\
$B_1$ & 0.3 & 0.52 & \cellcolor{backcolour} 0.21 & \cellcolor{backcolour} 0.25 & \cellcolor{backcolour} 0.04\\
$B_2$ & 0.2 & 0.40 & \cellcolor{backcolour} 0.16 & \cellcolor{backcolour} 0.24 & \cellcolor{backcolour} 0.08\\
$U$   & 1.6 & 0.92 & 0.44 & 0.56 & 0.12\\
\textit{Poisson} & & & & &\\
$B_1$ & 0.5 & 0.97 & \cellcolor{backcolour} 0.5 & \cellcolor{backcolour} 0.81 & \cellcolor{backcolour} 0.31\\
$B_2$ & 0.8 & 1.56 & \cellcolor{backcolour} 0.8 & \cellcolor{backcolour} 1.13 & \cellcolor{backcolour} 0.33\\
$U$   & 6.0 & 2.53 & 6.0 & 1.41 & -4.59\\
\bottomrule
\end{tabular}
\caption{\label{tab: var} 
Reconciliations that increase the  variance of the bottom variables (grey background).
For the Bernoulli case, we set $\phat_1 = 0.3$, $\phat_2 = 0.2$, and $\qH = [0.1, 0.2, 0.7]$, and after reconciliation we obtain $\ptil_1 = 0.52$, $\ptil_2 = 0.40$, and $\qT \approx [0.32, 0.44, 0.24]$.
}
\end{table}

We set $\phat_1 = 0.3$, $\phat_2 = 0.2$, and $\qH = [0.1, 0.2, 0.7]$,
which induce large incoherence.
The probability of coherence is $p_c = 0.17$, computed as:
\begin{equation}\label{eq: p_c coherence}
p_c = \sum_{\substack{u, b_1, b_2: \\ u = b_1 + b_2}} \pihat(u, b_1, b_2).
\end{equation} 
In this  case,  the variance of all variables increases after reconciliation
(Table~\ref{tab: var}).

\paragraph*{Poisson base forecast}

We now consider the case of Poisson independent base forecasts:
\[\Bhat_1 \sim \Poi\big(\lambdahat_1\big), \quad \Bhat_2 \sim \Poi\big(\lambdahat_2\big), \quad \Uhat \sim \Poi\big(\lambdahat_u\big),\]
with $\lambdahat_1,\lambdahat_2, \lambdahat_u > 0$.
We obtain a small probability of coherence by setting $\lambdahat_1 = 0.5$, $\lambdahat_2 = 0.8$, and $\lambdahat_u = 6.0$, which results in  $p_c = 0.03$.
We perform reconciliation via importance sampling (Section~\ref{sec: sampling from the rec distr}). 
The variance of the bottom variables, computed using samples, increases after reconciliation (Table~\ref{tab: var}).

\subsection{Reconciled mean in the non-Gaussian case}
\label{sec: rec mean}
In some cases, the reconciled mean of the upper variable can be lower than \textit{both} the mean of its base forecast and the  bottom-up mean. We call  this the ``concordant-shift effect''.
This happens  when reconciliation decreases the variance of 
base forecasts that are right-skewed,  which is often the case with count variables:
the right tail is shortened, lowering the expected values.

\begin{figure}[!h]
     \centering
     \begin{subfigure}[b]{0.45\textwidth}
         \centering
         \includegraphics[width=\textwidth]{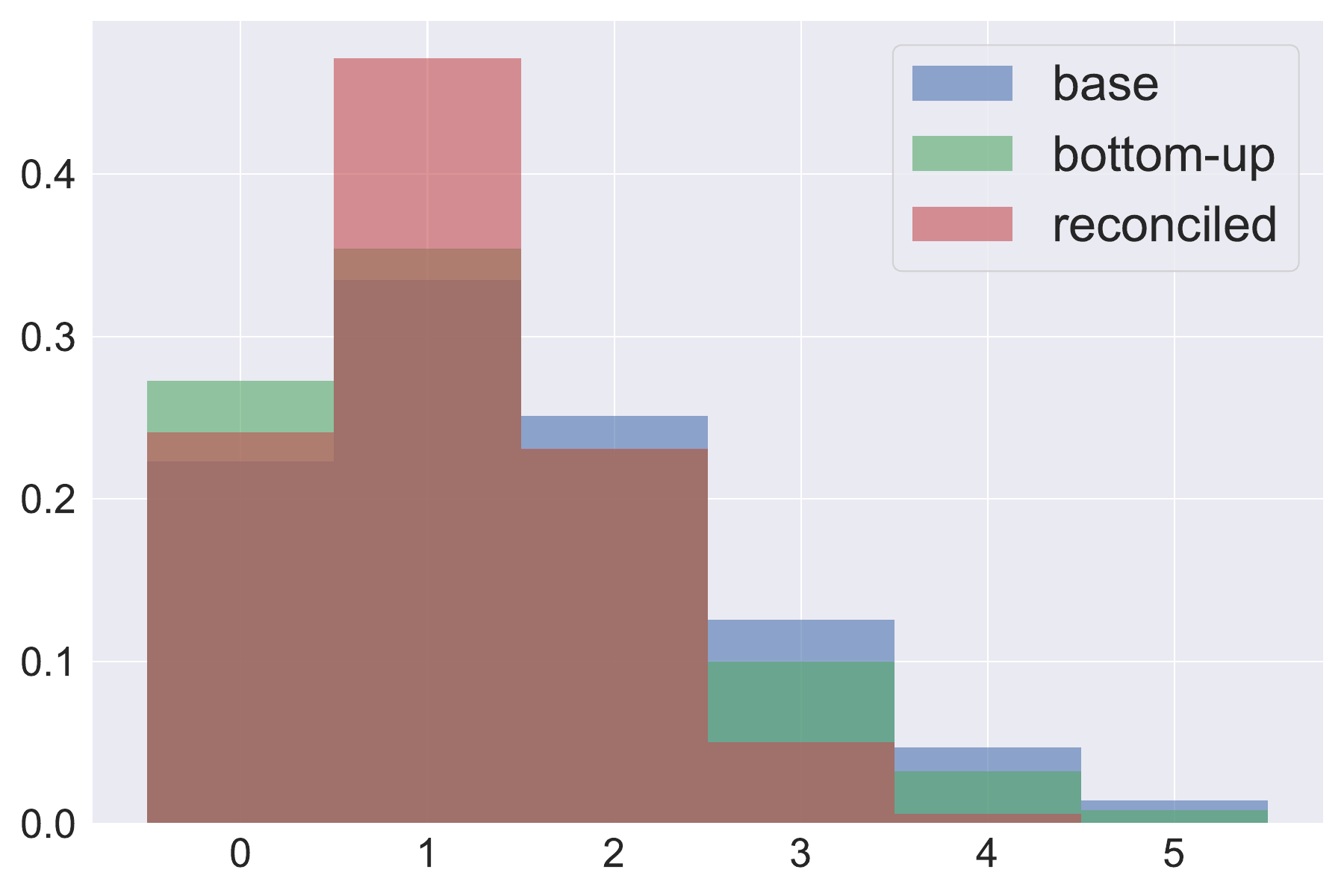}
         \caption{Concordant-shift effect.}
         \label{fig: poisson 0.5 0.8 1.5}
     \end{subfigure}
     \hfill
     \begin{subfigure}[b]{0.45\textwidth}
         \centering
         \includegraphics[width=\textwidth]{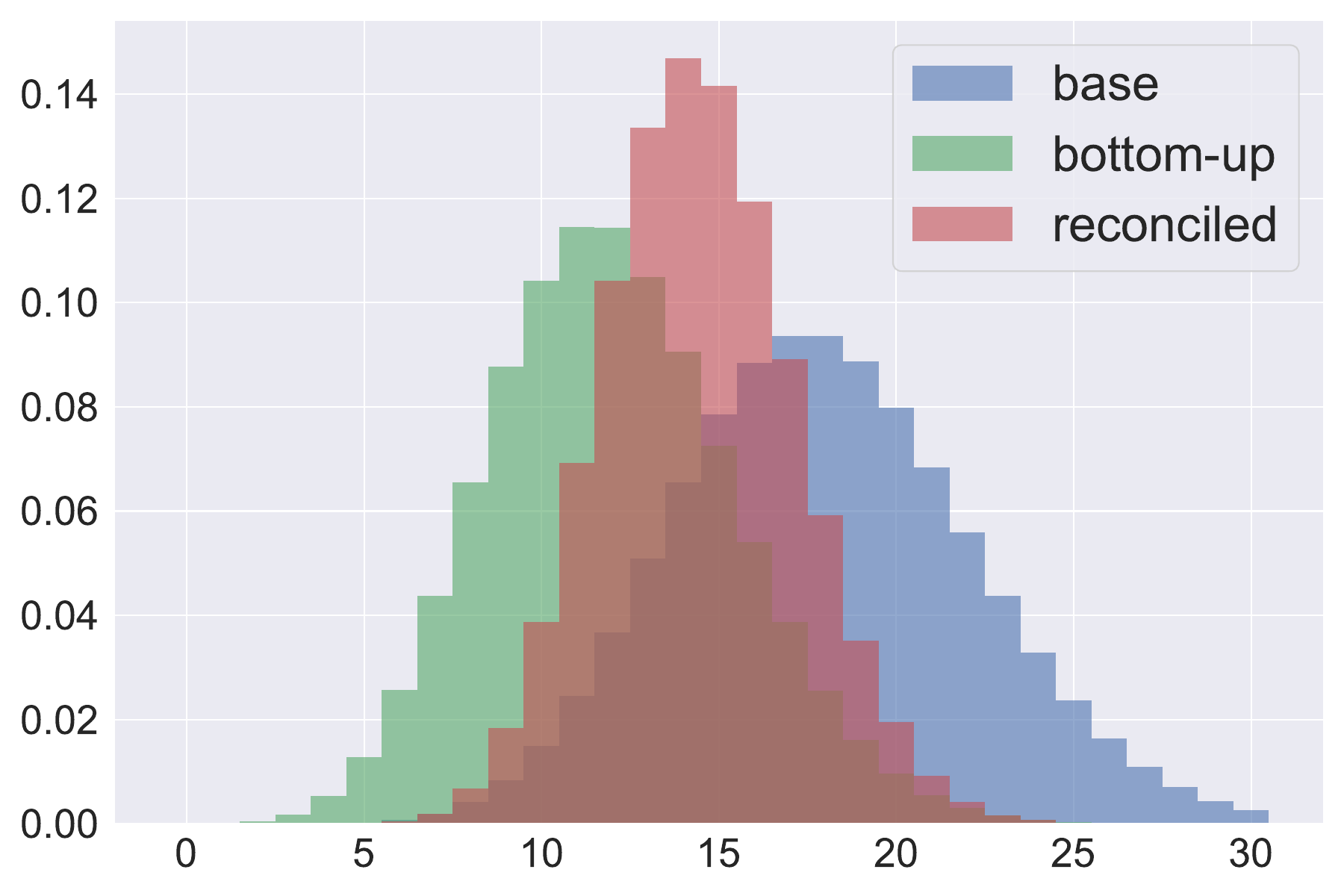}
         \caption{Combination effect.}
         \label{fig: poisson 5 7 18}
     \end{subfigure}
        \caption{Base, bottom-up, and reconciled  probability mass functions of $U$ in case of Poisson base forecasts.
        (a) $\lambdahat_1 = 0.5$, $\lambdahat_2 = 0.8$, $\lambdahat_u = 1.5$
        (b) $\lambdahat_1 = 5$, $\lambdahat_2 = 7$, $\lambdahat_u = 18$}
        \label{fig: poisson}
\end{figure}

To show the concordant-shift effect on the minimal hierarchy (Fig.~\ref{fig: simple hier gauss}),
we consider independent Poisson base forecasts with  $\lambdahat_1 = 0.5$, $\lambdahat_2 = 0.8$, and $\lambdahat_u = 1.5$. 
Thus, all base  forecasts convey information of low counts.
Reconciliation fuses  the base forecasts 
emphasizing the tendency towards 0: the mean of all the variables decreases after reconciliation (Table~\ref{tab: mean poisson}).
The reconciled distribution of the upper variable has a lower mean than both the base and bottom-up distributions (Fig.~\ref{fig: poisson 0.5 0.8 1.5}).
In contrast, we can induce the combination effect (Fig.~\ref{fig: poisson 5 7 18})  by
setting $\lambdahat_1 = 5$, $\lambdahat_2 = 7$, and $\lambdahat_u = 18$:
in this case, the base forecasts are less skewed and $p_c$ is smaller (Table~\ref{tab: mean poisson}).

\begin{table}[!ht]
    \centering
    \small
\begin{tabular}{l @{\hskip 8mm} c c c @{\hskip 8mm} c c c}
\toprule
& \multicolumn{3}{c @{\hskip 8mm}}{\textbf{concordant-shift effect}} & \multicolumn{3}{c}{\textbf{combination effect}} \\
& \multicolumn{3}{c @{\hskip 8mm}}{$p_c = 0.25$} & \multicolumn{3}{c}{$p_c = 0.04$} \\
\toprule
& \textit{base mean} & \textit{rec. mean} & $\Delta$ & \textit{base mean} & \textit{rec. mean} & $\Delta$ \\
\midrule
$B_1$ & 0.50 & 0.43 & \cellcolor{backcolour} -0.07 & 5.00 & 6.02 & \cellcolor{backcolour} 1.02 \\
$B_2$ & 0.80 & 0.68 & \cellcolor{backcolour} -0.12 & 7.00 & 8.43 & \cellcolor{backcolour} 1.43 \\
$U$ & 1.50 & 1.11 & \cellcolor{backcolour} -0.39 & 18.0 & 14.44 & \cellcolor{backcolour} -3.56 \\
\bottomrule
\end{tabular}
\caption{\label{tab: mean poisson} Mean before and after reconciliation; the base forecasts are Poisson.}
\end{table}

\section{Case study: modeling extreme market events}
\label{sec:experiments}
Credit default swaps (CDS) are financial instruments that guarantee insurance against the possible default of a given company (called ``reference company'') to the buyer. The CDS price is a function of the probability of default estimated by the market for that company. Thus, a  high value of the CDS spread corresponds to an increase in the risk of a company default. 

Following  \cite{raunig2011value},  an \textit{extreme market event} takes place when the value of the CDS spread on a given day exceeds the 90-th percentile of its distribution in the last trading year. In particular, we  forecast the extreme market events for the companies of the Euro Stoxx 50 index (\url{https://www.stoxx.com/}) in the  period 2005-2018, which includes  
3508 trading days.
As in \cite{agosto2020tree}, we consider the 29 
companies included in the index  having a regularly quoted CDS  and we divide them into five economic sectors: Financial (FIN), Information and Communication Technology (ICT), Manufacturing (MFG), Energy (ENG), and Trade (TRD). 
Following \cite{agosto2022multivariate}, we start from the CDS spread  time series retrieved from Bloomberg and we count the daily number of extreme events  for each sector, obtaining five daily time series.
The series include  3508 data points each; they  have low counts and a high frequency of zeros (Table~\ref{tab:descr}); they are all intermittent, with large  average inter-demand interval (ADI).  

\begin{table}[!htp] 
	\centering
	\small
	\begin{tabular}{lcccc}
		\toprule
		Sector             & Time series       & Mean         & Proportion of zeros & ADI\\
		\midrule
		FIN                & 10                        & 0.96       & 0.73 & 3.8\\
        
        \rowcolor{backcolour}
		ICT                & 4                         & 0.38       & 0.79 & 4.7\\
		
        MFG                & 7                         & 0.67       & 0.73 & 3.7\\
        
        \rowcolor{backcolour}
		ENG                & 5	                       & 0.48       & 0.80 & 4.9\\
		
        TRD                & 3                         & 0.29       & 0.81 & 5.2\\
		\bottomrule
	\end{tabular}
	\caption{Main characteristics of the  count time series.  A time series is considered intermittent if its ADI is \textgreater 1.32 \citep{SyntetosBoylan2005}.
 \label{tab:descr}}
\end{table}

\paragraph{Base forecasts and hierarchical structure}

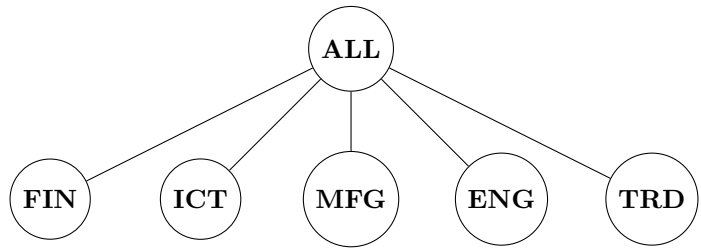
\begin{figure}[!h]
    \centering
\begin{tikzpicture} 
  \foreach \x/\label in {1/\textbf{FIN}, 2/\textbf{ICT}, 3/\textbf{MFG}, 4/\textbf{ENG}, 5/\textbf{TRD}} {
    \node[circle, draw] (bottom\x) at (2*\x,0) {\label};
  }
  
  \node[circle, draw] (top) at (6,2) {\textbf{ALL}};
  
  \foreach \x in {1,2,3,4,5} {
    \draw (top) -- (bottom\x);
  }
\end{tikzpicture}
 \caption{\label{fig: finTS hier} Hierarchical structure of the extreme events time series.}
\end{figure}

We organize the time series into a hierarchy with 5 bottom and 1 upper time series:
the five economic sectors constitute the bottom level, while their sum constitutes the  top level (Fig.~\ref{fig: finTS hier}). 

We compute the base forecasts for the bottom time series (counts in the different sectors) using the model of \cite{agosto2022multivariate}, whose 
predictive distribution is a multivariate negative binomial with a static vector of dispersion parameters  and a time-varying vector of location parameters following a score-driven dynamics \citep{Harvey_2013}. The model of \cite{agosto2022multivariate} extends to the multivariate case the modeling approach of \cite{agosto2016}, who proposed a Poisson autoregressive model with exogenous covariates (PARX) to measure systemic risk in the corporate default dynamics. According to \cite{agosto2022multivariate}, the predictive distribution computed at time $t$ for the count of time $t+1$ in sector $i$ is:
\begin{equation} \label{nb}
\yhat_{i,t+1} \sim NB(\mu_{i,t+1},\alpha_i),
\end{equation}
where $\mubf_t$ is the $k \times 1$ vector of location parameters, $k$ is the number of sectors, and $\alpha_i \geq 0$.
The model assumes the following dynamics:
\begin{equation} \label{score_driv}
\log(\mubf_{t+1})= \Cbf + \Dbf \log{\mubf_t} + \Ebf \dfrac{\ybf_t-\mubf_t}{\alphabf^T \mubf_t+1},
\end{equation}
where $\Cbf$ is a $k \times 1$ vector, while $\Dbf$ and $\Ebf$ are $k \times k$ matrices (see \cite{agosto2022multivariate} for detailed properties and estimation details).
Thus, the predicted event count in a given sector is a function of the past expectations ($\mubf_t$) and forecast errors ($\ybf_t-\mubf_t$) in the same sector and in the other sectors.  
The base forecasts for the upper time series are computed by fitting a univariate version of the model \citep{blasques2018} on the aggregate count time series.

The base forecasts for the bottom series measure financial risks at the sector level, accounting for the dependencies between sectors, through a shock propagation mechanism, besides an autoregressive component. The base upper forecasts express instead a measure of aggregate systemic financial risk.

\subsection{Experimental procedure}
As in \cite{agosto2022multivariate}, we estimate the model parameters  through in-sample maximum likelihood.
We then compute the 1-day ahead base forecasts for time $t+1$ by conditioning the model on the counts observed up to time $t$. 

We reconcile the in-sample base forecasts using importance sampling (Sect.~\ref{sec: sampling from the rec distr}). We perform 3508 reconciliations, drawing each time  $N=100,000$ samples from the reconciled distribution; each reconciliation is almost instantaneous (\textless 0.1 sec).

As recommended by \cite{panagiotelis2023probabilistic}, 
we compare base and reconciled distributions through the energy score (ES, \cite{szekely2013energy}):
\[
ES(P_t,\ybf_t) = \E_{P_t}\left[\|\ybf_t-\sbf_t\|^{\beta}\right] 
- \frac{1}{2} \E_{P_t}\left[\|\sbf_t-\sbf_t'\|^{\beta}\right],
\]
where $P_t$ is the forecast distribution on the whole hierarchy, $\sbf_t, \sbf_t'$ are a pair of independent random variables distributed as $P_t$, and $\ybf_t$ is the vector of the actual values of all the time series at time $t$.
We compute the ES, with $\beta = 2$, using the sampling approach of \cite{wickramasuriya2023probabilistic}.
We compute the ES of the joint predictive distribution of the upper and bottom time series; we thus have  a single ES for the entire hierarchy.

Moreover, we evaluate the prediction intervals using  the interval score (IS, \cite{gneiting2007strictly}):
\[
\text{IS}_\alpha(l_t,u_t;y_t) = (u_t - l_t) + \frac{2}{\alpha} (l_t - y_t) \mathbb{1}(y_t<l_t) + \frac{2}{\alpha} (y_t-u_t) \mathbb{1}(y_t>u_t),
\]
where $\alpha \in (0,1)$, $l_t$ and $u_t$ are the lower and upper bounds of the $(1-\alpha) \times 100 \,\%$  prediction interval and $y_t$ is the actual value of the time series at time $t$. We use $\alpha = 0.1$, i.e. we score prediction interval whose nominal coverage is 90\%.

Finally, we evaluate the point forecasts measuring the squared error (SE)
and the absolute error (AE):
\begin{align*}
\text{SE} &= \left( y_t - \hat{y}_{t\mid t-1}\right)^2,\\
\text{AE} &= \left| y_t - \hat{y}_{t\mid t-1}\right|,
\end{align*}
where $\hat{y}_{t\mid t-1}$ denotes the optimal point forecast.
The optimal point forecast depends on the error measure:
it is the median of the predictive distribution for AE, and the expected value for SE \citep{kolassa2016evaluating, kolassa2020best}.

\paragraph{Skill score}
The \textit{skill score} measures the improvement of the reconciled forecasts over the  base  forecasts.
For example, the skill score for AE is defined as:
\begin{equation} \label{eq: def skill score}
\text{skill}(\textit{reconc}, \textit{base}) 
= \frac{\text{AE}(\textit{base}) - \text{AE}(\textit{reconc})}
{\left(\text{AE}(\textit{base}) + \text{AE}(\textit{reconc})\right) / 2}.
\end{equation}
For all indicators, a positive  skill score implies  an improvement 
of the reconciled forecast compared to the base forecasts, and vice versa.
The skill score defined in \eqref{eq: def skill score} is symmetric, allowing  to fairly  compare base and reconciled forecasts. 
For instance, a skill score of 1 implies that the loss function has been reduced by three times: $(3-1) / ((3+1)/2) = 2/2 = 1$. Analogously, a skill score of $-1$ implies a three-fold worsening of the loss function.
Moreover, the skill score is bounded between $-2$ and $2$.
\cite{difonzo2022forecast} compare competing approaches by computing the geometric mean of the indicators. However,  this is not suitable for count time series, where the SE and the AE are often zero.

\begin{table}[!ht]
    \centering
    \small
\begin{tabular}{c b c b c b c}
\toprule
   & \textit{ALL} & \textit{FIN} & \textit{ICT} & \textit{MFG} & \textit{ENG} & \textit{TRD} \\
\midrule
\textbf{ES} & \multicolumn{6}{c}{0.89}  \\
\textbf{IS} & 0.87 & 1.15 & 0.20 & 1.07 & 0.22 & 0.18 \\
\textbf{SE} & 0.82 & 1.10 & 1.11 & 1.07 & 1.12 & 1.11 \\
\textbf{AE} & -0.02 & -0.02 & -0.02 & -0.02 & -0.02 & -0.02 \\
\bottomrule
\end{tabular}
\caption{\label{tab: skill scores} Average skill scores for  the different  time series. Positive values indicate an improvement of the reconciled forecasts over the base forecasts. The ES is computed with respect to the joint distribution on the entire hierarchy.}
\end{table}

\begin{figure}[h!]
    \centering
    \includegraphics[width=0.65\textwidth]{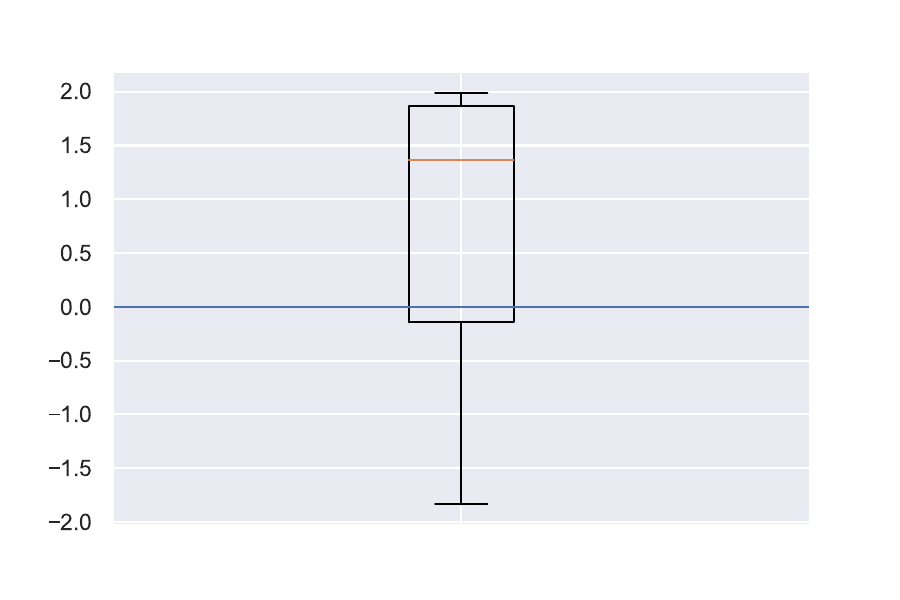}
    \caption{\label{fig: boxplot ss es} Boxplot of the skill scores on ES, over 3508 reconciliations.}
\end{figure}

\begin{table}[!htp]
    \centering
    \small
\begin{tabular}{cbcbcbc}
\toprule
   & \textit{ALL} & \textit{FIN} & \textit{ICT} & \textit{MFG} & \textit{ENG} & \textit{TRD} \\
\midrule
\textit{base} & 6.9 & 3.2 & 1.1 & 1.9 & 1.3 & 0.9 \\
\textit{reconc.} & 3.3 & 2.1 & 0.9 & 1.2 & 1.0 & 0.8 \\
\bottomrule
\end{tabular}
    \caption{Width of the $90\%$ prediction interval, averaged over 3508 reconciliations.}
    \label{tab: width intervals average} 

\bigskip

\begin{tabular}{cbcbcbc}
\toprule
   & \textit{ALL} & \textit{FIN} & \textit{ICT} & \textit{MFG} & \textit{ENG} & \textit{TRD} \\
\midrule
\textit{base} & 96\% & 98\% & 98\% & 98\% & 98\% & 99\% \\
\textit{reconc.} & 91\% & 95\% & 97\% & 97\% & 97\% & 98\% \\
\bottomrule
\end{tabular}
\caption{Coverage of the $90\%$ prediction intervals, assessed on 3508 reconciliations.}
\label{tab: coverage intervals}
\end{table}

In Table~\ref{tab: skill scores} we report the skill scores averaged over the 3508 reconciliations.
Reconciliation largely improves the ES (with an average skill score of $0.89$) and the 
IS  (average skill score ranging between 0.2 and 1.1, depending on the chosen time series).
The boxplot of the skill scores on ES on each day (Fig.~\ref{fig: boxplot ss es}) confirms the improvement due to reconciliation. 
As a further insight, reconciliation  reduces by 15-50\% the width of the prediction intervals  (Table~\ref{tab: width intervals average}) without compromising their coverage  (Table~\ref{tab: coverage intervals}).
We observe large skill scores (0.8-1) on the SE.
On the other hand, the skill core on AE is close to 0.
Indeed, often the median of the predictive distribution is already  0,
and it does not change after reconciliation.

Hence, in our experiments, reconciliation yields a major improvement over  the base forecasts, confirming  the positive effect of 
probabilistic reconciliation
\citep{corani2023,zambon2022} and
point forecast reconciliation \citep{KOURENTZES-elucidate}
for intermittent time series.

\paragraph{Coherent vs optimal point forecast for the upper time series}

Even if we have a reconciled distribution,
only the SE-optimal point forecasts are  coherent.
The AE-optimal point forecasts, i.e., the medians of the reconciled distributions, are instead generally incoherent \citep{kolassa22we}.

Hence, besides optimal point forecasts, we evaluate the coherent point forecasts: we  take the sum of the medians of the reconciled bottom distributions and use it as upper point forecast.
We compute the mean absolute error for the upper time series obtained using the AE-optimal, the coherent, and the base point forecasts (Table~\ref{tab: AE ALL}). 
As expected, the AE worsens by imposing coherence rather than optimality; yet, it remains better than the AE of the base forecasts.

\begin{table}[!ht]
    \centering
    \small
\begin{tabular}{l c c c}
\toprule
   & \textit{optimal} & \textit{coherent} & \textit{base} \\
\midrule
\textbf{ALL} & 1.10 & 1.24 & 1.33 \\
\bottomrule
\end{tabular}
\caption{\label{tab: AE ALL} Mean absolute error for the upper time series ``ALL''.}
\end{table}

\subsection{Analysis of the reconciled mean and variance}

\begin{figure}[!htp]
    \centering
    \begin{subfigure}{1.\textwidth}
        \includegraphics[width=\linewidth]{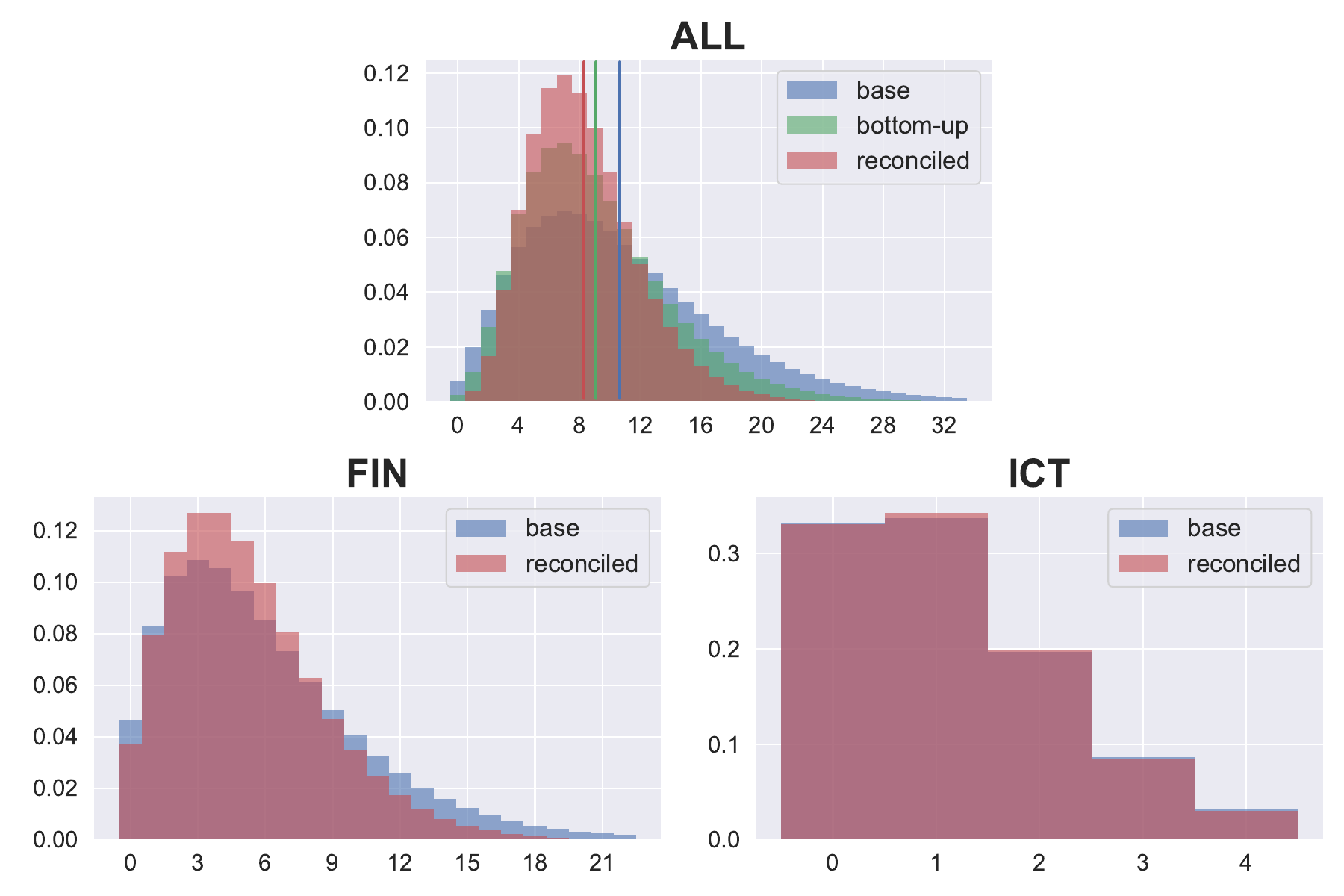}
        \subcaption{Concordant-shift effect. The reduction of variance of the asymmetric distribution shortens the right tail and decreases the mean.}
        \label{fig: pmf 123}
    \end{subfigure}

\bigskip
\bigskip

    \begin{subfigure}{1.\textwidth}
        \includegraphics[width=\linewidth]{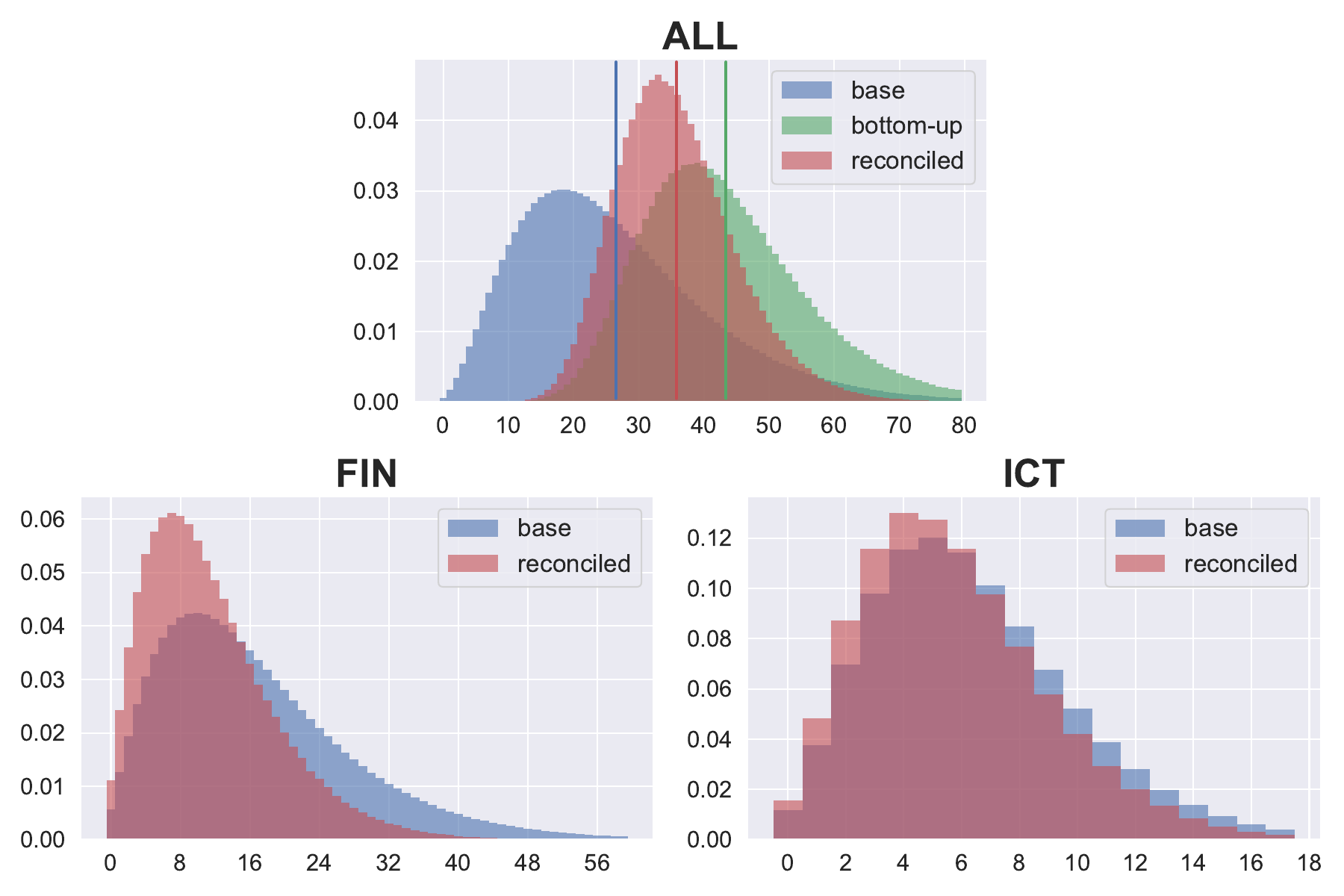}
        \subcaption{Combination effect}
        \label{fig: pmf 1699}
    \end{subfigure}
    \caption{Concordant-shift vs combination effect. We plot the base and reconciled probability mass functions for  ``ALL'', ``FIN'', and ``ICT''. For  ``ALL'', we also show the bottom-up pmf. The vertical lines correspond to the means of the distributions.}
\end{figure}

\begin{figure}[!h]
    \centering
    \begin{subfigure}{1.\textwidth}
        \includegraphics[width=\linewidth]{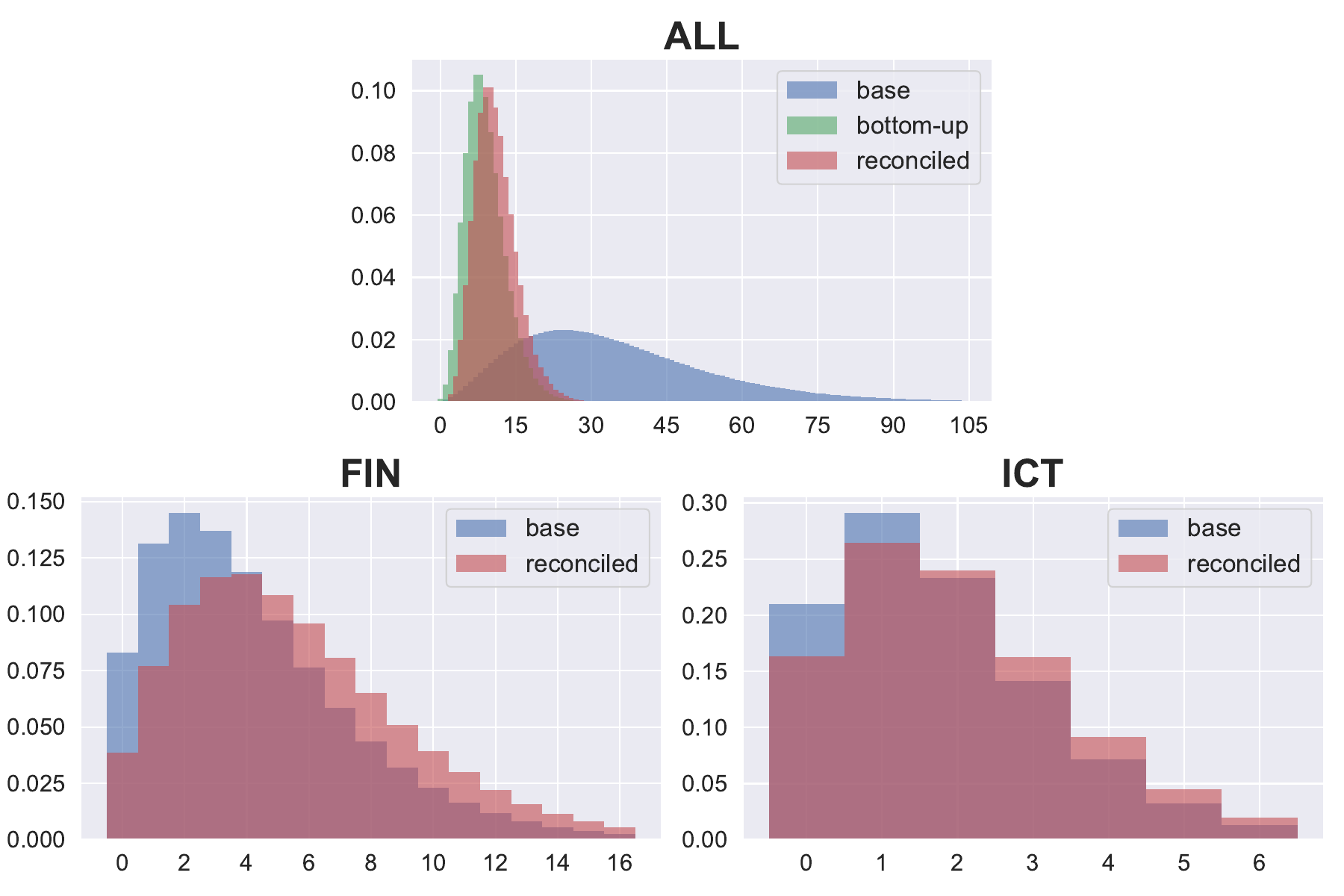}
    \end{subfigure}
    \caption[]{An example of reconciliation which increases the variance of the bottom forecasts, due to a large incoherence of the base forecasts (note the $X$-axis of ``ALL'').
    Variance of ``FIN'': $11.4$ (base), $14.3$ (reconciled).
    Variance of ``ICT'': $2.2$ (base), $2.6$ (reconciled).}
    \label{fig: pmf 2307}
\end{figure}

In most reconciliations (96\%), we observe the concordant-shift effect, i.e., the reconciled upper mean is lower than both the bottom-up and the base upper mean.
In Fig.~\ref{fig: pmf 123}, we show an example.
Reconciliation largely reduces the variance
of ``ALL'' and ``FIN'',    shortening the right tail of the distribution.
Within the bottom time series, reconciliation mostly affects FIN, which is characterized by the largest counts and overdispersion.
The predictive distribution for other time series, such as  ICT (shown in figure), are less affected by  reconciliation, being characterized by lower counts and variability.
The reduction of the variance  shifts  the expected values towards zero, since the base distributions have a positive skewness.

In Fig.~\ref{fig: pmf 1699}, we show an example of the combination effect.
The reconciled distribution of ``ALL'' is a combination between its  base forecast and the probabilistic bottom-up.  
Reconciliation decreases the expected values of the bottom time series, while increasing the  expected value of the upper time series. 

The variance of all the variables decreases in most cases (97\%).
However, in Fig.~\ref{fig: pmf 2307} we show an example in which  the variance of the  bottom variables increases after reconciliation because of large incoherence of the  base forecasts.
The forecast of the upper time series, which is characterized by large uncertainty, is instead 
sharply shifted towards smaller values.

\section{Conclusions}
\label{sec: conclusions}
We  proved that reconciliation via conditioning decreases the variance of all variables in the Gaussian framework, while it can increase or decrease the variance of discrete variables, depending on the incoherence of the base forecasts.
We also discussed two different effects that the reconciliation can have on the upper mean.
The empirical analysis of time series of extreme market events confirmed our theoretical insights.
We leave as future research the study of the theoretical properties of the reconciliation of other continuous  non-Gaussian distributions, including skewed ones. 

\section{Acknowledgements}

Work partially funded by the Swiss National Science Foundation (grant 200021\_212164/1), by the Hasler foundation (project 23057), and by the European Union's Horizon 2020 research and innovation programme (grant 101016233).

\bibliographystyle{elsarticle-harv}
\bibliography{biblio}

\newpage
\appendix

\section{Reconciled distribution in the Gaussian case}
\label{appendix: derivation gaussian}

\paragraph{Bottom distribution}

Let us define $\Tbf \in \rr^{n \times n}$ as
\begin{equation*}
\Tbf = \begin{bmatrix} \textbf{0} & \Ibf_m \\
\Ibf_{n-m} & -\Abf \end{bmatrix},
\end{equation*}
and let $\Zbf := \Tbf\YH$.
Hence, $\Zbf$ is Gaussian:
\begin{equation}
\Zbf \sim \mathcal{N}\left( \Tbf \yH, \,\Tbf\SigH_Y \Tbf^T \right).
\end{equation}
We have
\begin{align}\label{eq: T Sigma T}
\Tbf \yH &= \begin{bmatrix} \bH \\ \uH - \Abf \bH \end{bmatrix}, \nonumber \\
\Tbf \SigH_Y \Tbf^T &= \begin{bmatrix} \SigH_B & \SigH_{UB}^T - \SigH_B \Abf^T \\
\SigH_{UB} - \Abf \SigH_B & \Qbf \end{bmatrix}, \end{align}
where $\Qbf= \SigH_U - \SigH_{UB} \Abf^T - \Abf \SigH_{UB}^T + \Abf \SigH_B \Abf^T$.
Since 
\[\Zbf = \begin{bmatrix} \BH \\ \UH - \Abf\BH \end{bmatrix} =: \begin{bmatrix} \Zbf_1 \\ \Zbf_2 \end{bmatrix},\]
the reconciled bottom distribution is given by the conditional distribution of $\Zbf_1$ given $\Zbf_2=0$. 
Since the covariance matrix $\SigH_Y$ is assumed to be positive definite, the covariance matrix $\Qbf$ of $\Zbf_2$ is also positive definite; hence, we have
\begin{equation*}
\Zbf_1 \,|\, \Zbf_2 = 0 \sim \mathcal{N}\left(\bT, \,\SigT_B \right), 
\end{equation*}
where
\begin{align}
\bT &= \bH + \left(\SigH_{UB}^T - \SigH_B \Abf^T\right) \Qbf^{-1} (\Abf \bH - \uH), \label{eq: reconciled gaussian mean bottom appendix} \\
\SigT_B &= \SigH_B - \left(\SigH_{UB}^T - \SigH_B \Abf^T\right) \Qbf^{-1} \left(\SigH_{UB}^T - \SigH_B \Abf^T\right)^T. \label{eq: reconciled gaussian variance bottom appendix}
\end{align}

\paragraph{Upper distribution}

Since $\UT = \Abf \BT$, we have that $\UT \sim \mathcal{N}\left( \uT, \,\SigT_U \right)$, with
\begin{align} \label{eq: gauss rec upper mean and var}
\uT = \Abf \bT, \qquad
\SigT_U = \Abf \SigT_B \Abf^T.
\end{align}
If we define $\Dbf := \SigH_U - \SigH_{UB} \Abf^T$, from \eqref{eq: gauss rec upper mean and var} and \eqref{eq: reconciled gaussian mean bottom appendix} we have
\begin{align}
\uT &= \Abf \bH + \left(\Abf \SigH_{UB}^T - \Abf \SigH_B \Abf^T\right) \Qbf^{-1} (\Abf \bH - \uH) \nonumber \\
&= \Abf \bH + \left(\Dbf - \Qbf\right) \Qbf^{-1} (\Abf \bH - \uH) \nonumber \\
&= \Abf \bH + \Dbf \Qbf^{-1} (\Abf \bH - \uH) - (\Abf \bH - \uH) \nonumber \\
&= \uH + \left(\SigH_U - \SigH_{UB} \Abf^T\right) \Qbf^{-1} (\Abf \bH - \uH). \nonumber
\end{align}
Moreover, from \eqref{eq: gauss rec upper mean and var} and \eqref{eq: reconciled gaussian variance bottom appendix}:
\begin{align}
\SigT_U &= \Abf \SigH_B \Abf^T - \left(\Abf \SigH_{UB}^T - \Abf \SigH_B \Abf^T\right) \Qbf^{-1} \left(\Abf \SigH_{UB}^T - \Abf \SigH_B \Abf^T\right)^T  \nonumber \\
&=  \Abf \SigH_B \Abf^T - \left(\Dbf - \Qbf\right) \Qbf^{-1} \left(\Dbf^T - \Qbf\right)  \nonumber \\
&=  \Abf \SigH_B \Abf^T - \Dbf \Qbf^{-1} \Dbf^T + \Dbf + \Dbf^T - \Qbf \nonumber \\
&= \SigH_U - \left(\SigH_U - \SigH_{UB} \Abf^T\right) \Qbf^{-1} \left(\SigH_U - \SigH_{UB} \Abf^T\right)^T. \nonumber
\end{align}

\paragraph{Relation to \cite{corani_reconc}}

In \cite{corani_reconc}, the reconciliation framework was defined is the following way:
\begin{itemize}
    \item $\Bbf' \sim \mathcal{N}\big(\bH, \SigH_B\big)$,
    \item $\UH' = \Abf \Bbf' + \epsbf^u$,
\end{itemize}
where the error term $\epsbf^u$ was assumed to have zero mean and to be jointly Gaussian with $\Bbf'$.
The reconciled distribution on the bottom time series was then obtained via a Bayes' update by conditioning on $\UH' = \uH$, the point forecast for the upper time series.

In this paper, we adopt a slightly different framework. 
We assume that the base forecasts are jointly Gaussian:
\[
\YH = \begin{bmatrix} \UH \\ \BH \end{bmatrix}
\sim \mathcal{N}\left( \yH, \,\SigH_Y \right),
\]
hence we have 
$\BH \sim \mathcal{N}\big(\bH, \SigH_B\big)$
and 
$\UH \sim \mathcal{N}\big(\uH, \SigH_U\big)$.
The reconciled distribution on the bottom time series is obtained by conditioning on $\UH = \Abf \BH$.
This is equivalent to the framework of \cite{corani_reconc}, if we set $\BH = \Bbf'$ and $\UH = \uH - \epsbf^u$.
The hypothesis of joint normality of $(\Bbf',\, \epsbf^u)$ is thus replaced by the hypothesis of joint normality of  $(\BH,\, \UH)$.
Indeed, \eqref{eq: reconciled gaussian mean bottom appendix} and \eqref{eq: reconciled gaussian variance bottom appendix} are precisely the formulae (11) and (12) in \cite{corani_reconc}, if we set
\begin{equation}\label{eq: equivalence cov matr}
\SigH_{UB} = - k_h \Mbf_1^T, \quad
\SigH_B = k_h \SigH_{B,1}, \quad
\SigH_U = k_h \SigH_{U,1}.
\end{equation}
Note that in this paper we only deal with one step ahead forecasts (i.e., $h=1$), hence $k_h = 1$ and can be omitted.
Since the covariance matrices correspond to those of \cite{corani_reconc}, as shown in \eqref{eq: equivalence cov matr}, we conclude that $\SigH_Y$ can be estimated as the covariance matrix of the residuals of the entire hierarchy. 

\section{Proof of Proposition \ref{prop: var gauss}}
\label{app: proof var ineq}

First, $\Qbf$ is positive definite, as it is the covariance matrix of $\UH - \Abf\BH$ (see \ref{appendix: derivation gaussian}).
Hence, $\Qbf^{-1}$ is also positive definite, and the matrices
\begin{align}
\Gbf := \left(\SigH_{UB}^T - \SigH_B \Abf^T\right) \Qbf^{-1} \left(\SigH_{UB}^T - \SigH_B \Abf^T\right)^T, \nonumber \\
\Hbf := \left(\SigH_U - \SigH_{UB} \Abf^T\right) \Qbf^{-1} \left(\SigH_U - \SigH_{UB} \Abf^T\right)^T, \nonumber
\end{align}
are positive semi-definite. 

From \eqref{eq: reconciled gaussian variance bottom}, we have that, for each $i=1,\dots,m$
\[\Var(\Btil_i) = \Var(\Bhat_i) - G_{ii} \le \Var(\Bhat_i),\]
as $G_{ii} \ge 0$ since the matrix $\Gbf$ is positive semi-definite.
Analogously, we have 
\[\Var(\Util_j) = \Var(\Uhat_j) - H_{jj} \le \Var(\Uhat_j),\]
for all $j=1,\dots,n-m$.

\section{Proof of Proposition \ref{prop: reconc variance}}
\label{app: proof rec var}

Let us denote $Z := \mathbb{1}_{\{ \UH = \Abf \BH \}}$, so that $Z=1$ when the constraint is satisfied, and $0$ otherwise.
By the law of total variance \citep{Weiss2005ACI}, for any $j=1,\dots,m$, we have
\begin{equation}
\label{eq: law of total variance}
\Var\left(\Bhat_j\right) = 
\E_Z\left[\Var\big(\Bhat_j|Z\big)\right] + 
\Var_Z\left(\E\big[\Bhat_j|Z\big]\right).
\end{equation}
Since 
\begin{equation*}
\E\big[\Bhat_j | Z\big] = \begin{cases}
\E\big[\Bhat_j | \UH = \Abf \BH\big] \quad \text{if } Z = 1 \\
\E\big[\Bhat_j | \UH \neq \Abf \BH\big] \quad \text{if } Z = 0,
\end{cases}
\end{equation*}
we have that $\E[\Bhat_j | Z] = a + (b-a) \, Z$,
where $a := \E[\Bhat_j | \UH \neq \Abf \BH]$ and $b := \E[\Bhat_j | \UH = \Abf \BH]$.
Note that $Z \sim \text{Bernoulli}(p_c)$, with
$p_c := \prob(\UH = \Abf \BH)$.
Hence
\begin{align}
\Var_Z\left(\E\big[\Bhat_j|Z\big]\right)
&= \Var_Z\big(a + (b-a) \, Z\big) \nonumber \\
&= (b-a)^2 p_c (1-p_c). \label{eq: var(E)}
\end{align}
Moreover, since 
\begin{equation*}
\Var\big[\Bhat_j | Z\big] = \begin{cases}
\Var\big[\Bhat_j | \UH = \Abf \BH\big] \quad \text{if } Z = 1 \\
\Var\big[\Bhat_j | \UH \neq \Abf \BH\big] \quad \text{if } Z = 0,
\end{cases}
\end{equation*}
we have 
\begin{equation}
\label{eq: E(var)}
\E_Z\left[\Var\big(\Bhat_j|Z\big)\right] = p_c \, \Var\big[\Bhat_j | \UH = \Abf \BH\big] 
+ (1-p_c) \, \Var\big[\Bhat_j | \UH \neq \Abf \BH\big].    
\end{equation}
From \eqref{eq: law of total variance}, \eqref{eq: var(E)}, and \eqref{eq: E(var)}, we obtain
\begin{align*}
\Var\big(\Bhat_j\big) &= p_c \, \Var\big[\Bhat_j | \UH = \Abf \BH\big] + (1-p_c) \, \Var\big[\Bhat_j | \UH \neq \Abf \BH\big] \nonumber \\
& \, + p_c (1-p_c) \, (a-b)^2,
\end{align*}
from which 
\begin{align*}
\Var\big[\Btil_j\big]
&= \Var\big[\Bhat_j | \UH = \Abf \BH\big] \\
&= \frac{\Var\big(\Bhat_j\big) - (1-p_c) \, \Var\big[\Bhat_j | \UH \neq \Abf \BH \big] - p_c (1-p_c) \, (a-b)^2}{p_c}.
\end{align*}

\section{Bernoulli example}
\label{app: example bernoulli}

Let us denote by $\pihat_1$ and $\pihat_2$ the probability mass functions of $\Bhat_1$ and $\Bhat_2$, so that $\pihat_1(0) = 1 - \phat_1$, $\pihat_1(1) = \phat_1$, and $\pihat_1(k)=0$ for any $k \neq 0,1$.
The probability mass function $\pihat_U$ of $\Uhat$ is defined as $\pihat_U(0) = \qhat_0$, $\pihat_U(1) = \qhat_1$, $\pihat_U(2) = \qhat_2$, and $\pihat_U(k) = 0$ for any $k \neq 0,1,2$.

Since, from \eqref{eq: reconciled distribution}, the reconciled probability mass function $\pitil_B$ is given by 
\begin{equation*}
\pitil_B(b_1,b_2) \propto \pihat_{1}(b_1) \pihat_{2}(b_2) \pihat_U(b_1+b_2),
\end{equation*}
the reconciled bottom distribution can be expressed as
\begin{equation*}
(\widetilde{B}_1, \widetilde{B}_2) = \begin{cases}
(0,0) \qquad \text{prob} = (1-\phat_1)(1-\phat_2)\qhat_0 \, / \, p_c \\
(1,0) \qquad \text{prob} = \phat_1(1-\phat_2)\qhat_1 \, / \, p_c \\
(0,1) \qquad \text{prob} = (1-\phat_1)\phat_2\qhat_1 \, / \, p_c \\
(1,1) \qquad \text{prob} = \phat_1\phat_2\qhat_2 \, / \, p_c,
\end{cases}
\end{equation*}
where the normalizing constant $p_c := (1-\phat_1)(1-\phat_2)\qhat_0 + \phat_1(1-\phat_2)\qhat_1 + (1-\phat_1)\phat_2\qhat_1 + \phat_1\phat_2\qhat_2$ is the 
probability of coherence, defined in \eqref{eq: p_c coherence}.
Hence
\begin{equation*}
\widetilde{B}_1 \sim \textit{Bernoulli}\,(\ptil_1), \quad
\widetilde{B}_2 \sim \textit{Bernoulli}\,(\ptil_2),
\end{equation*}
with
\begin{align}
\ptil_1 &= \frac{[(1-\phat_2)\qhat_1 + \phat_2 \qhat_2] \phat_1}
{S}, \nonumber \\
\ptil_2 &= \frac{[(1-\phat_1)\qhat_1 + \phat_1 \qhat_2] \phat_2}
{S}. \label{eq: ptil bernoulli}
\end{align}
Finally,
\begin{equation}
\label{eq: qtil bernoulli}
\Util = \begin{cases}
0 \qquad \text{prob} = (1-\phat_1)(1-\phat_2)\qhat_0 / S \\
1 \qquad \text{prob} = (\phat_1+\phat_2-2\phat_1\phat_2)\qhat_1 / S \\
2 \qquad \text{prob} = \phat_1\phat_2\qhat_2 / S.
\end{cases}
\end{equation}

\end{document}